\documentclass[aps,prl,onecolumn,12pt,superscriptaddress,notitlepage]{revtex4}

\pdfoutput=1

\usepackage{amsmath}
\usepackage[pdftex]{graphicx}

\newcommand{\erw}[1]{\langle #1 \rangle}

\begin{document}

\title{Impact of ultrafast electronic damage in single particle x-ray imaging experiments}
\date{\today}

\author{U.~Lorenz}
	\affiliation{DESY, Notkestrasse 85, D-22607 Hamburg, Germany}
\author{N.M.~Kabachnik}
	\affiliation{Skobeltsyn Institute of Nuclear Physics, Lomonosov Moscow State University, 119991 Moscow, Russia}
	\affiliation{European XFEL GmbH, Albert-Einstein-Ring 19, D-22761 Hamburg, Germany}
\author{E.~Weckert}
\affiliation{DESY, Notkestrasse 85, D-22607 Hamburg, Germany}
\author{I.A.~Vartanyants}
\email[Corresponding author: ]{ivan.vartaniants@desy.de}
\affiliation{DESY, Notkestrasse 85, D-22607 Hamburg, Germany}
\affiliation{National Research Nuclear University, ``MEPhI'', 115409 Moscow, Russia}

\begin{abstract}
In single particle coherent x-ray diffraction imaging experiments, performed at x-ray free-electron lasers (XFELs), samples are exposed to intense x-ray pulses to obtain single-shot diffraction patterns.
The high intensity induces electronic dynamics on the femtosecond time scale in the system, which can reduce the contrast of the obtained diffraction patterns and adds an isotropic background.
We quantify the degradation of the diffraction pattern from ultrafast electronic damage by performing simulations on a biological sample exposed to x-ray pulses with different parameters.
We find that the contrast is substantially reduced and the background is considerably strong only if almost all electrons are removed from their parent atoms. This happens at fluences of at least one order of magnitude larger than provided at currently available XFEL sources.
\end{abstract}

\pacs{61.80.Cb,87.53.Ay,87.64.Bx}

\maketitle

Modern crystallography has developed efficient tools for protein structure determination up to atomic resolution provided that sufficiently large high-quality crystals are available \cite{crystallography1,crystallography2}.
Presently there are two major limitations for the future development of this approach, namely crystallization and radiation damage. It is well known that the majority of protein macromolecules, especially membrane proteins, do not crystalize.
At the same time, radiation damage at conventional x-ray sources limits the available resolution of non-crystalline biological samples to few tens of nanometers \cite{Damage}.

To break this radiation barrier limit in the structure determination of non-crystalline samples it was suggested \cite{Neutze00} to use high power x-ray free-electron lasers (XFELs) \cite{LCLS,SCSS,XFEL} in so-called 'single-molecule' experiments \cite{GaffneyChapman,imagingStuff}.
In this technique, single biological particles are injected in the strong coherent XFEL beam in a random orientation, and their diffraction patterns in the far field are measured.
By collecting a large series of such diffraction patterns from a reproducible sample and applying orientation determination techniques \cite{orientation1,orientation2,orientation3}, the full three-dimensional (3D) diffraction pattern in reciprocal space can be assembled.
Using phase-retrieval algorithms \cite{PhaseRetrieval1,PhaseRetrieval2} a 3D image of the electron density of the particle can be reconstructed in the final step.
With XFEL pulses shorter than few tens of femtoseconds, no radiation damage of the sample due to Coulomb explosion is expected during the pulse propagation \cite{Neutze00}.
In this case the diffraction pattern should contain, at least in principle, information about the sample in its initial undamaged state.
This method has been successfully applied recently to protein nanocrystals \cite{NanoCrystals} and viruses \cite{MimiVirus} with sizes of few hundred nanometers.
However, the resolution of non-crystalline samples was found to be far below the expected sub-nanometer limit.

To increase the resolution and to allow the imaging of smaller samples, ideally individual proteins, the photon fluence should be increased by several orders of magnitude.
In this high-power regime, however, atoms are known to undergo strong ionization \cite{Rohringer,Young}; we can thus expect that the electronic density will be altered considerably during the scattering process.
This raises the question how much the diffraction pattern is affected by this electronic damage, and if there is an upper limit to the fluence, beyond which the quality of the diffraction pattern decreases substantially and prevents a successful reconstruction.
This problem was first addressed in Ref. \cite{Quiney}, which was focused on adapting phase-retrieval algorithms to incorporate the effect of electronic damage.
In this letter, we quantitatively analyse the degradation of the diffraction patterns for a wide range of incoming x-ray pulse parameters.

In a 'single-particle' experiment, the time-integrated intensity at the detector can be expressed as
\begin{gather}
\label{intro::initial}
	I(\mathbf{q}) = \int j(t) |A(\mathbf{q}, t) |^2 \mathrm{d}t
	\ ,
\end{gather}
where the scattering amplitude per unit flux is given in the kinematical approximation by $A(\mathbf{q}, t) = \sum_i f_i(q, t) \exp(\imath \mathbf{qR_i}(t))$.
Here, $\mathbf{q}$ is the scattering vector, $j(t)$ is the intensity of the incoming pulse, and $f_i(q,t)$, $\mathbf{R}_i(t)$ are the time-dependent form factor and position vector of the $i$-th atom, respectively.
In the following we approximate the atomic form factors to be spherically symmetric and consider only small samples with a size of tens of nanometers. We also assume a homogenous and coherent illumination over the sample area, and neglect retardation effects \cite{Hau-Riege08} in Eq.~\eqref{intro::initial}. Note that the recorded diffraction pattern described by Eq. \eqref{intro::initial} has the form of an \emph{incoherent} sum over instantaneous \emph{coherent} diffraction patterns $|A(\mathbf{q}, t)|^2$. Clearly, under certain experimental conditions, this summation can decrease the contrast of the measured diffraction pattern.

The radiation damage induced by intense XFEL pulses is basically a two-step process.
In a first step, the sample is strongly ionized on a time scale of few femtoseconds \cite{Hau-Riege04,Young,Sang-Kil,shakeoff,Berrah10}.
The ejected electrons leave the sample, which thereby accumulates a space charge \cite{Hau-Riege12}.
In a second step, the system disintegrates through a Coulomb explosion within tens to hundreds of femtoseconds \cite{Hau-Riege04,Neutze00}.
This results in a motion of the atoms from their equilibrium positions, $\mathbf{R}_i = \mathbf{R}_i(t)$, which seriously degrades the diffraction pattern through the incoherent summation in Eq. \eqref{intro::initial}.
The use of pulses with a duration of few femtoseconds should prevent the degradation from the Coulomb explosion \cite{Neutze00}, but the \emph{electronic} damage cannot be avoided.
Obviously, the ionization lowers the atomic form factors, which reduces the scattered signal and thus the achievable resolution.
In addition, it introduces an explicit time-dependence in the form factors, which can also reduce the contrast of the diffraction pattern.

For our analysis, we assume that a large number of single-shot patterns have been measured, aligned, and averaged perfectly.
Since we are interested only in electronic damage, we also require that the atomic positions do not change during the pulse propagation, and that the sample is perfectly reproducible for all shots.
With these assumptions, an average of Eq. \eqref{intro::initial} over many pulses gives
\begin{gather}
\label{theory::averaging}
	\erw{I(\mathbf{q})} = I_0 \sum_{i,j} F_{ij}(q)
		\mathrm{e}^{\imath \mathbf{q}(\mathbf{R}_j - \mathbf{R}_i)}
	\ .
\end{gather}
Here, $I_0 = \int j(t) \mathrm{d}t$ is the total fluence of the x-ray pulse, the brackets $\langle \ldots \rangle$ denote averaging over many pulses and the summation is performed over all atoms in the sample.
A normalized form factor matrix $\hat{\mathbf{F}}$ with matrix elements
\begin{gather}
\label{theory::matrix}
	F_{ij}(q) = \frac{1}{I_0}
		\int j(t) \erw{f_i^{\ast}(q,t) f_j(q,t)} \mathrm{d}t
\end{gather}
is introduced in Eq. \eqref{theory::averaging}.
We assume here that all FEL pulses are identical \cite{SASE_note}, so that the intensity $j(t)$ can be taken out of the average in Eq. \eqref{theory::matrix}.
Note that the time integration in Eq. \eqref{intro::initial} that causes the degradation of contrast is now encapsulated in the form factor matrix $\hat{\mathbf{F}}$.

To proceed with the analysis, the time-dependent form factors can be written as a sum $f_i(q,t) = \erw{f_i(q,t)} + \delta f_i(q,t)$ with average values $\erw{f_i(q,t)}$ and pulse-to-pulse fluctuations $\delta f_i(q,t)$ (by definition $\erw{\delta f_i(q,t)} = 0$).
We can then express the average of the form factor product in Eq.~\eqref{theory::matrix} as $\erw{f_i^{\ast}(q,t) f_j(q,t)} = \erw{f_i^{\ast}(q,t)} \erw{f_j(q,t)} + \erw{\delta f_i^{\ast}(q,t) \delta f_j(q,t)}$.
If we finally assume that the ionization of atoms at different positions $i,j$ is uncorrelated \cite{Hau-Riege04}, the second term reduces to $\erw{\delta f_i^{\ast}(q,t) \delta f_j(q,t)} = \delta_{ij} \erw{|\delta f_i(q,t)|^2}$.
The matrix $\hat{\mathbf{F}}$ then naturally decomposes into two contributions,
\begin{align}
\label{theory::separation}
	F_{ij}(q) &= W_{ij}(q) + \delta_{ij} B_{i}(q)
	\ ,
\intertext{where $\hat{\mathbf{W}}(q)$ is an Hermitian matrix \cite{Quiney_note}}
\label{theory::matrix_A}
	W_{ij}(q) &= \frac{1}{I_0} \int j(t) \erw{f_i^{\ast}(q, t)} \erw{f_j(q,t)} \mathrm{d}t
	\ ,
\intertext{and the vector $\mathbf{B}(q)$ is defined as}
\label{theory::vector_B}
	B_i(q) &= \frac{1}{I_0} \int j(t)
		\Big\langle |\delta f_i(q,t)|^2 \Big\rangle \mathrm{d}t
\ .
\end{align}
Substituting Eq.~\eqref{theory::separation} in Eq.~\eqref{theory::averaging} we obtain for the average intensity
\begin{gather}
\label{theory::averaging2}
	\erw{I(\mathbf{q})} = I_{\text{W}}(\mathbf{q}) + I_{\text{B}}(q)
	\ ,
\end{gather}
where $I_{\text{W}}(\mathbf{q}) = I_0 \sum_{i,j} W_{ij} \exp [i\bf{q}(\bf{R}_j-\bf{R}_i)]$ and $I_{\text{B}}(q) = I_0 \sum_i B_i(q)$.
Our derivation shows that the average intensity $\erw{I(\mathbf{q})}$ in a 'single-particle' experiment can be presented as a sum of two contributions.
The structural term, $I_{\text{W}}(\mathbf{q})$, is determined by the average form factor values $\erw{f_i(q,t)}$, and the background term, $I_{\text{B}}(q)$, by their fluctuations $\erw{|\delta f_i(q,t)|^2}$.

\begin{figure}
	\includegraphics[width=\linewidth]{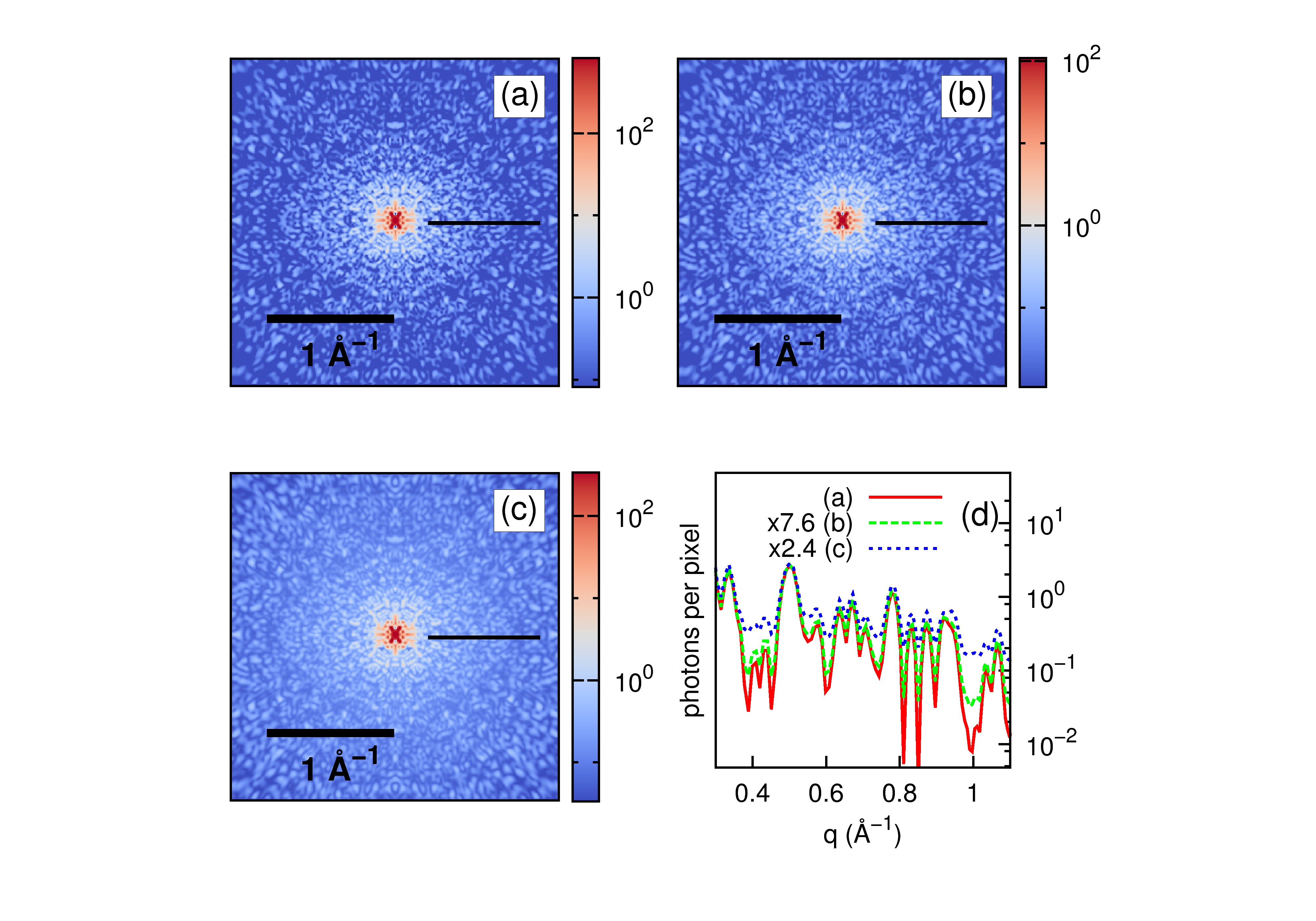}
	\caption{\label{fig::Sample}(Color online) Averaged two-dimensional
		diffraction patterns of the biological sample in logscale for an x-ray pulse
		of 5 fs FWHM and 3.1 keV photon energy.
		(a) Diffraction pattern of the neutral sample without electronic damage at a fluence of 10$^{14}$
		photons/$\mu$m$^2$.
		(b, c) Diffraction patterns of the ionized sample at a fluence of 10$^{14}$ photons/$\mu$m$^2$ (b) and
		 10$^{16}$ photons/$\mu$m$^2$ (c).
		(d) Intensity profiles for (a-c) along the black lines.
        The intensities in (d) have been rescaled to be equal at $q=0$.
	}
\end{figure}

The structural term $I_{\text{W}}(\mathbf{q})$ in Eq.~\eqref{theory::averaging2} can be presented as a sum of coherent modes \cite{Quiney}. To show that, we formally decompose the matrix $\hat{\mathbf{W}}(q)$ as \cite{Hermitian_note}
\begin{gather}
\label{theory::modes}
	W_{ij}(q) = \sum_{\alpha} c_{\alpha}(q) \ v_i^{(\alpha)\ast}(q) v_j^{(\alpha)}(q)
	\ ,
\end{gather}
where $c_{\alpha}, \mathbf{v}^{(\alpha)}$ are the eigenvalues and normalized eigenvectors of $\hat{\mathbf{W}}(q)$.
With this decomposition, we have for $I_{\text{W}}(\mathbf{q})$ in Eq.~\eqref{theory::averaging2}
%
%
\begin{align}
\label{theory::averaged_explicit}
	I_{\text{W}}(\mathbf{q}) &= I_0 \sum_{\alpha} c_{\alpha}(q) |A^{(\alpha)}(\mathbf{q})|^2
	\ ,
\end{align}
where $A^{(\alpha)}(\mathbf{q}) = \sum_{i} v_i^{(\alpha)}(q) \exp(\imath \mathbf{q}\mathbf{R}_i)$.
An inspection of Eq. \eqref{theory::averaged_explicit} shows that the structural contribution $I_{\text{W}}(\mathbf{q})$ to the averaged intensity $\erw{I(\mathbf{q})}$ is described by a sum of coherent modes $|A^{(\alpha)}(\mathbf{q})|^2$. Each mode produces a diffraction pattern with full contrast with the weight given by the eigenvalue $c_{\alpha}(q)$. This summation reduces the contrast of the final diffraction pattern depending on the incoming pulse parameters.
%
%
%
The magnitude of the contribution of the modes can be described by the parameter \cite{degree_of_coherence_note}
\begin{gather}
\label{theory::coherence}
	\zeta(q) = \frac{\mathrm{Tr}(\hat{\mathbf{W}}(q)\hat{\mathbf{W}}(q)^{\dagger})}{(\mathrm{Tr}\hat{\mathbf{W}}(q))^2}
		= \frac{\sum_{\alpha} c_{\alpha}^2(q)}{(\sum_{\alpha} c_{\alpha}(q))^2}
	\ ,
\end{gather}
where $\mathrm{Tr}\hat{\mathbf{W}} = \sum_i W_{ii}$ denotes the trace operation.
By definition, $0 < \zeta(q) \leq 1$, and $\zeta(q)$ is equal to one only if there is a single mode at a certain value of $q$.

The background term $I_{\text{B}}(q)$ in Eq.~\eqref{theory::averaging2} gives rise to an additional isotropic background.
As a result of our assumption that the ionization of atoms at different positions is uncorrelated, this background does not contain any structural information and cannot be included in the mode decomposition described by Eq. \eqref{theory::averaged_explicit}.
%
%
To quantify the relative contribution of this background we introduce the ratio
\begin{gather}
\label{theory::backgroundContrib}
	\Gamma(q) = \frac{I_{\text{B}}(q)}{\langle I_{\text{W}}(\mathbf{q}) \rangle_{\phi}}
\end{gather}
between the background and the angular average of the structural term $I_{\text{W}}(\mathbf{q})$ in Eq.~\eqref{theory::averaging2}.
Here, the brackets $\langle \ldots \rangle_{\phi}$ denote angular averaging of the diffraction pattern.

To analyze the effect of the electronic damage, we have simulated \cite{moltrans} averaged diffraction patterns using Eq. \eqref{theory::averaging2} for a human adenovirus penton base chimera \cite{sample}.
The sample has a dodecahedral shape with a diameter of 27 nm and contains about 200,000 non-hydrogen atoms, giving a mass density of about 0.5 g/cm$^3$.
In our simulations we considered an incoming x-ray beam with a photon energy of 3.1 keV (corresponding to the wavelength $\lambda$ = 4 \AA{}), and 12.4 keV ($\lambda$ = 1 \AA{}) focused to a 100x100 nm$^2$ area at the sample position. The detector was positioned 100 mm downstream from the sample and had a size of 200$\times$200 mm$^2$ for the soft, and 400$\times$400 mm$^2$ for the hard x-ray energies. This provided a maximum scattering angle of 45$^{\circ}$ and available resolution $d_{\mathrm{max}} =$5.2 \AA{} for the soft, and 63.4$^{\circ}$ ($d_{\mathrm{max}} =$0.95 \AA{}) for the hard x-ray energies.
%
%

The averages of the form factors $\erw{f_i(q,t)}$ and their fluctuations $\erw{|\delta f_i(q,t)|^2}$ in Eqs.~\eqref{theory::matrix_A} and \eqref{theory::vector_B} were evaluated with a rate equation approach \cite{Sang-Kil}.
We included ionization from photoionization, Auger decay and electron impact ionization into the rate equations \cite{supplement}.
In our simulations, only carbon, nitrogen and oxygen atoms were considered.
The contribution from sulphur and hydrogen atoms can be safely neglected; the former appear only in trace quantities, while the latter hardly scatter.

Simulated diffraction patterns for an x-ray pulse with 5 fs FWHM, 3.1 keV photon energy and fluences of 10$^{14}$ and 10$^{16}$ photons/$\mu$m$^2$ are shown in Fig.~\ref{fig::Sample}.
We can immediately see that the electronic damage reduces the contrast of the diffraction pattern in comparison with the undamaged sample (Fig.~\ref{fig::Sample}(a)).
For the lower fluence (Fig.~\ref{fig::Sample}(b)), this mainly leads to a slight smearing of the deep minima of the diffraction pattern.
At the same time, for the higher fluence (Fig.~\ref{fig::Sample}(c)), most of the details of the diffraction pattern are smeared out.

\begin{figure}
	\includegraphics[width=\linewidth]{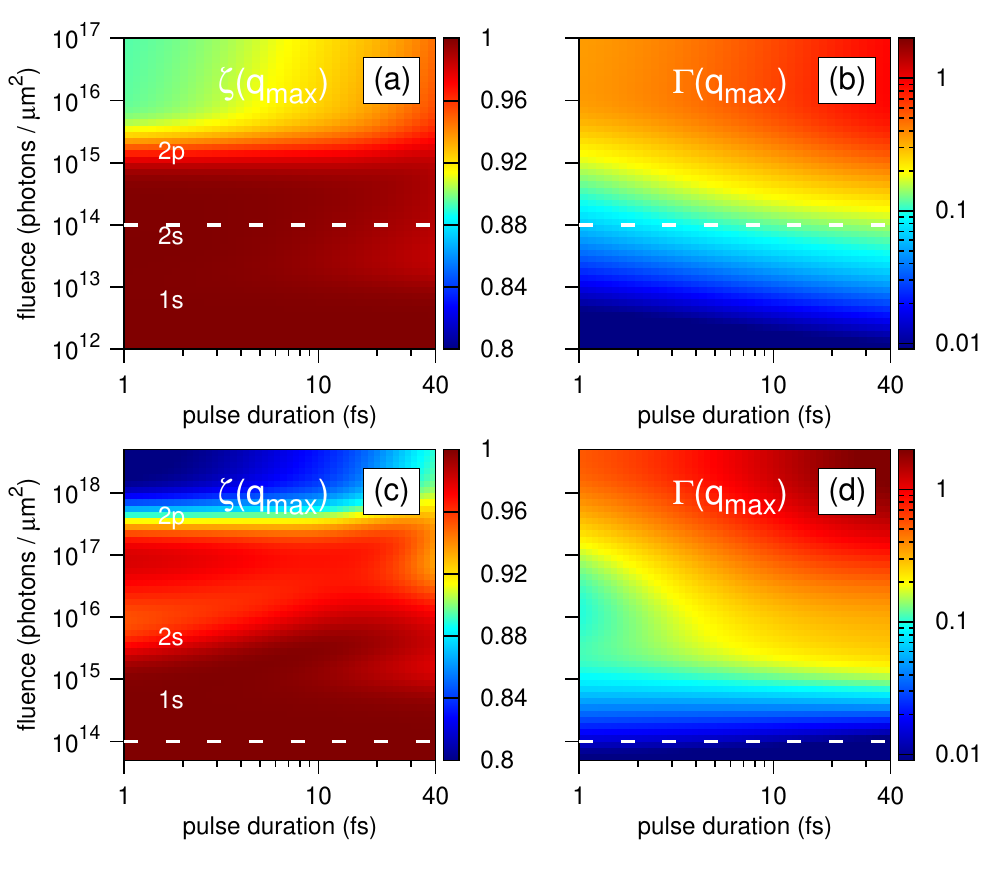}
	\caption{\label{fig::results}(Color online)
		(a, c) Contrast $\zeta(q_{\text{max}})$, and (b, d) relative background contribution $\Gamma(q_{\text{max}})$
        calculated at $q_{\text{max}}$ corresponding to a resolution of 5.2 \AA{} at 3.1 keV photon energy (a, b) and a resolution of 0.82 \AA{} at 12.4 keV photon energy (c, d).
		The labels indicate the approximate fluences for the saturation of the ionization of the respective shells:
        1s of neutral carbon, 2s of C$^{2+}(1s^02s^22p^2)$, and 2p of C$^{4+}(1s^02s^02p^2)$. The dashed line in all figures corresponds to the fluence of 10$^{14}$ photons/$\mu$m$^2$ presently achievable at XFELs.
	}
\end{figure}

To determine the relative contribution of the multimode decomposition and the background to the diffraction pattern, we have calculated the parameters $\zeta(q)$ and $\Gamma(q)$ [Eqs.~\eqref{theory::coherence} and \eqref{theory::backgroundContrib}] at the momentum transfer value $q_{\text{max}}=2\pi/d_{\text{max}}$ for a wide range of x-ray pulse parameters (see Fig.~\ref{fig::results}).
At the soft x-ray energy (Fig.~\ref{fig::results}(a)), $\zeta(q_{\text{max}})$ is close to one up to the fluences of $~10^{15}$ photons/$\mu$m$^2$ when all electrons are photoionized.
At higher fluences it rapidly drops down to a value of 0.9.
At hard x-ray energies (Fig.~\ref{fig::results}(c)), the contrast drops in two steps.
First, slightly (down to a value of 0.95), at the fluences from $10^{15}$ to $10^{16}$ photons/$\mu$m$^2$ when core electrons are efficiently ionized.
Then, strongly (to a value of 0.8), at extreme fluences higher than $10^{17}$ photons/$\mu$m$^2$ when all electrons are removed from the atoms.
It follows from our simulations that the contrast is substantially reduced only if almost all scattering electrons are removed from their parent atoms.
This will not cause any problems with the presently available fluences but will become a substantial factor at fluences at least one order of magnitude higher.

The dependence of the background contribution $\Gamma(q_{\text{max}})$ on the pulse parameters (Figs.~\ref{fig::results}(b,d)) is qualitatively similar for the soft and hard x-ray energies.
The background is negligible for the lowest fluences and rises continuously with increasing fluence.
It becomes significant (more than 10\%)  at fluences higher than $10^{14}$ ($10^{15}$) photons/$\mu$m$^2$ at the soft (hard) x-ray energies.
At high fluences, $\Gamma(q_{\text{max}})$ ranges from 0.4 to 0.9 at soft x-ray energies and from 0.5 to 2 for hard x-ray energies for pulse durations from 1 fs to 40 fs.
For hard x-ray energies and fluences of about $10^{16}$ photons/$\mu$m$^2$, the background is reduced for pulses shorter than $3$ fs. We attribute this to formation of hollow atoms \cite{Sang-Kil}, which reduces the form factor fluctuations due to the K-shell depletion.

\begin{figure}
 \includegraphics[width=\linewidth]{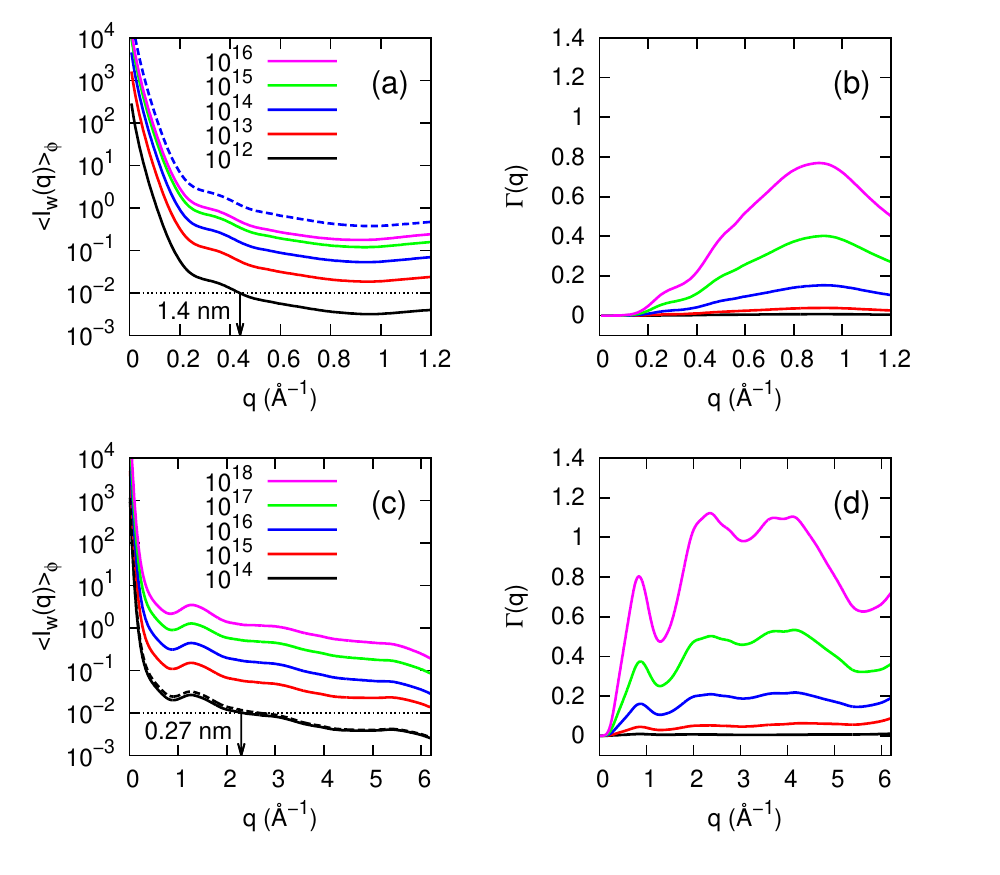}
 \caption{\label{fig::high-res}(Color online)
	(a, c) Angular averaged scattered signal $\left<I_{\text{W}}(\mathbf{q})\right>_{\phi}$,
	and (b, d) relative background contribution $\Gamma(q)$ for a 5 fs FWHM x-ray pulse
	at different fluences; (a, b) 3.1 keV, (c, d) 12.4 keV photon energy.
	Dashed curves correspond to the intensity scattered by a sample without electronic damage.
	Dotted lines in (a,c) denote the cut-off of $10^{-2}$ photons per
    Shannon angle, and the labels denote the achievable resolution.
	The curves have been smoothed with a Gaussian filter to remove high-frequency oscillations from the speckle pattern.
}
\end{figure}

As an important outcome of our simulations we could estimate an effect of electronic damage on the scattered signal  with increased fluence (see Fig.~\ref{fig::high-res}(a,c)).
For a soft x-ray pulse with $10^{14}$ photons/$\mu$m$^2$ the sample scatters one order of magnitude less photons than a corresponding undamaged sample.
In contrast, for high x-ray energies, both signals are on the same level (see Fig.~\ref{fig::high-res}(c)).
In the first case, the fluence is sufficient to remove most of the (core and valence) electrons from the atoms (see Fig.~\ref{fig::results}(a)), while for high x-ray energies the photoionization cross sections are substantially smaller (see Fig.~\ref{fig::results}(c)).
Furthermore, once the fluence increases beyond the level of $10^{13}$ ($10^{15}$) photons/$\mu$m$^2$ for soft (hard) x-rays, an increase of the scattered signal by \textit{one} order of magnitude requires an increase of the incoming intensity by \textit{three} orders of magnitude due to excessive ionization (compare with \cite{Sang-Kil}).

To estimate limitations to the achievable resolution introduced by the electronic damage we analyzed the $q-$dependence of the structural term $\left<I_{\text{W}}(q)\right>_{\phi}$ and the background ratio $\Gamma(q)$ (see Fig.~\ref{fig::high-res}).
We assumed that the resolution is determined by the criterion of measuring $10^{-2}$ photons per Shannon angle \cite{orientation1,orientation2}.
At hard x-ray energies, a high resolution below 1 \AA{} can be, in principle, achieved with a fluence of 10$^{15}$ photons/$\mu$m$^2$ (see Fig.~\ref{fig::high-res}(c)).
For these parameters, the background contribution $\Gamma(q)$ is about 5\%.
For higher fluences, however, the background quickly increases; at a fluence of $10^{16}$ photons/$\mu$m$^2$, the background signal is about 20\% of the scattered signal already at the resolution of 3 \AA{} (see Fig.~\ref{fig::high-res}(d)).
Similar behavior can be observed at soft x-ray energies at lower resolution and fluences (see Figs.~\ref{fig::high-res}(a,b)).This clearly demonstrates that electronic damage during propagation of ultrafast pulses limits the achievable resolution at high fluences.

In summary, we have shown that electronic damage can significantly reduce the contrast of single-particle diffraction patterns and gives rise to an isotropic background if high power FEL pulses are used.
We have derived quantities to describe these effects, and calculated them for a typical biological sample illuminated by an intense x-ray pulse. Our simulation have shown that at soft x-ray energies sub-nanometer resolution can be, in principle, achieved for fluences higher than 10$^{13}$ photons/$\mu$m$^2$, however, at fluences higher than 10$^{14}$ photons/$\mu$m$^2$ the background will give a substantial contribution to diffraction patterns. For hard x-ray energies, 2.7 \AA{} resolution is feasible for presently available XFEL sources. No contrast degradation or background contribution is expected in this case. Raising the fluence up to the limit of 10$^{15}$ photons/$\mu$m$^2$ will allow to reach 1 \AA{} resolution limit. However, further increase in the fluence will rapidly increase the background contribution to diffraction patterns. 


\begin{acknowledgments}
We acknowledge helpful discussions with A. Singer, O. Yefanov, and J. Gulden.
Part of this work was supported by BMBF Proposal 05K10CHG.
\end{acknowledgments}

\bibliography{references}

\begin{thebibliography}{37}
\bibitem[{SAS()}]{SASE_note}
\bibinfo{note}{Present self-amplified spontaneous emission (SASE) XFELs have
  typically few percent variation in pulse to pulse intensity, spectrum and
  microbunch structure. However, as shown in Refs. \cite{Rohringer,Young} this
  does not affect the ionization dynamics substantially. Self-seeded XFELs
  \cite{Saldin} are expected to produce high-power stable pulses}.
\bibitem[{Her()}]{Hermitian_note}
\bibinfo{note}{Note that this decomposition is formally equivalent to the
  coherent-mode decomposition of partially coherent fields
  \cite{MandelWolf,Andrej}. This can be seen directly by evaluating the mutual
  coherence function on a grid, $W_{ij} \equiv W(\mathbf{r}_i, \mathbf{r}_j)$
  and replacing integrals by corresponding summations. Similar to this case,
  Eq. \eqref{theory::modes} follows directly from the fact that
  $\hat{\mathbf{W}}$ is Hermitian and positive semidefinite, which also implies
  that the eigenvalues $c_{\alpha}$ are real-valued, non-negative numbers.}
\bibitem[{deg()}]{degree_of_coherence_note}
\bibinfo{note}{This quantity is identical to the normalized degree of coherence
  to quantify the global coherence of a partially coherent source
  \cite{Andrej}.}
\bibitem[{mol()}]{moltrans}
\bibinfo{note}{All diffraction patterns were calculated using the computer code
  MOLTRANS.}
\end{thebibliography}

\newpage

\appendix
\section*{Supplement: Implementation of the rate equations}

Here, we present the technical details of the implementation of the rate equations.

The rate equations are based on the approach developed in \cite{Sang-Kil}.
We first define a set of states that each atom type can occupy.
As such states, we consider the energetically lowest states for all different occupations of the electronic shells.
For example, carbon can have between zero and two electrons in each of the 1s, 2s, 2p shells, yielding a total of 27 distinct states.
We can then calculate the time-dependent occupation probability of the $\xi$-th state, $p_\xi(t)$, by solving a set of coupled differential equations of the form
\begin{gather}
	\label{appendixB::rate_equation}
	\dot p_\xi(t) = \sum_{\eta\neq \xi} R_{\xi\eta}(t) p_{\eta}(t) - R_{\eta\xi}(t) p_{\xi}(t)
	\ .
\end{gather}
Here, $R_{\xi\eta}(t)$ denotes the total time-dependent rate of transition from state $\eta$ to state $\xi$.
As initial condition, we start with all atoms in the neutral ground state.

In our simulations, the total rate was composed of four parts,
\begin{gather}
	\hat{\mathbf{R}}(t) = \hat{\mathbf{R}}^{\text{photo}}(t) + \hat{\mathbf{R}}^{\text{Auger}}
		+ \hat{\mathbf{R}}^{\text{escape}}(t) + \hat{\mathbf{R}}^{\text{trap}}(t)
	\ ,
\end{gather}
where $\hat{\mathbf{R}}^{\text{photo}}$ is the rate of photoionization, $\hat{\mathbf{R}}^{\text{Auger}}$ is the Auger rate, and $\hat{\mathbf{R}}^{\text{escape}}$, $\hat{\mathbf{R}}^{\text{trap}}$ are the rates of secondary ionisation from escaping and trapped electrons, respectively.

The photoionization rate is calculated from $\hat{\mathbf{R}}^{\text{photo}}(t) = \hat{\boldsymbol{\sigma}}^{\text{photo}} j(t)$.
The photoionization cross sections $\hat{\boldsymbol{\sigma}}^{\text{photo}}$ as well as the Auger rates $\hat{\mathbf{R}}^{\text{Auger}}$ are calculated within the Hartree-Fock-Slater (HFS) approximation \cite{HermanSkillman}; the explicit equations can be found for example in \cite{Sang-Kil}.

It is known that secondary ionization can significantly change the ionization behaviour of the atoms \cite{Hau-Riege04}, so we also accounted for these effects.
We use a different treatment for photoelectrons on one hand, and Auger and secondary electrons on the other hand.

Photoelectrons are assumed to originate in the center of the particle, and leave it instantaneously.
We consider the particle as a homogenous sphere.
With these approximations, the rate of secondary electrons produced by the photoelectrons is given by
\begin{gather}
	\hat{\mathbf{R}}^{\text{escape}}(t) = \hat{\boldsymbol{\sigma}}^{\text{impact}}(\hbar \omega_0)
		\frac{R}{V} \ \frac{\mathrm{d}N_{\text{photo}}}{\mathrm{d}t}
	\ ,
\end{gather}
where $R, V$ are the radius and volume of the molecule, and $\text{d}N_{\text{photo}} / \text{d}t$ is the flux of produced photoelectrons.
The impact cross sections $\hat{\boldsymbol{\sigma}}^{\text{impact}}$ were calculated using the Binary-Encounter-Bethe model \cite{NIST} with electron orbital parameters obtained from the HFS calculations.
When a secondary electron is produced, its kinetic energy is set to a constant value $E_0$, which we chose to be 25 eV \cite{Hau-Riege04}.

Auger electrons and all secondary electrons are assumed to be trapped by the charged particle, and to thermalize instantaneously into a homogenous electron gas with a Maxwell-Boltzmann distribution of velocities.
The temperature of the gas is given by the average kinetic energy $\overline{E}$ of the trapped electrons as $\overline{E} = 3/2 kT$.
The rate of collisional ionization is then calculated as
\begin{gather*}
	\hat{\mathbf{R}}^{\text{trap}}(t) = \erw{\hat{\boldsymbol{\sigma}}^{\text{impact}}(v) v}_T \ n_{\text{trap}}(t)
	\ .
\end{gather*}
Here, $n_{\text{trap}}$ is the density of trapped electrons, $v$ is the velocity of the particles, and $\erw{}_T$ is the average with the appropriate Boltzmann factors.

The electron density and total kinetic energy of the trapped electron gas are calculated from the set of equations
\begin{align}
	\frac{\mathrm{d}n_{\text{trap}}}{\mathrm{d}t} &= \sum_{\alpha} \sum_{\xi\eta}
		\biggl( R^{\text{Auger}}_{\xi\eta; \alpha} + R^{\text{escape}}_{\xi\eta;\alpha}(t)
		+ R^{\text{trap}}_{\xi\eta;\alpha}(t) \biggr)p_{\eta;\alpha}(t) \varrho_{\alpha}\\
	\frac{\text{d}E_{\text{trap}}}{\text{d}t} &=
		 \sum_{\alpha} \sum_{\xi\eta} \biggl(
		R^{\text{Auger}}_{\xi\eta; \alpha} E_{\xi\eta;\alpha}^{\text{Auger}}
		+ R^{\text{escape}}_{\xi\eta;,\alpha}(t) E_0
		- R^{\text{trap}}_{\xi\eta;\alpha}(t) E_{\xi\eta;\alpha}^{\text{coll}} \biggr)
		p_{\eta;\alpha}(t) \varrho_{\alpha} V
	\ .
\end{align}
The index $\alpha$ denotes the atom type (carbon, nitrogen etc.) with corresponding atom density $\varrho_{\alpha}$.
The energies $E_{\xi\eta;\alpha}^{\text{Auger}}, E_{\xi\eta;\alpha}^{\text{coll}}$ and $E_0$ are the energy of the released Auger electrons, the binding energy of the electron that is released through collisional ionization, and the energy assigned to secondary electrons produced by photoelectrons, respectively.
We should note that this model is a simplified version of the model used by Hau-Riege et al. \cite{Hau-Riege04}, and was chosen for its simplicity.
However, it is able to reproduce most features of the more sophisticated simulations in \cite{Hau-Riege04}, giving some confidence that it does capture the essential physics.

Solving the differential equations \eqref{appendixB::rate_equation} yields the time-dependent occupation probabilities $p_{\xi_i}(t)$ for each state $\xi_i$ of atom $i$.
Furthermore, from the HFS calculations, we can also obtain the form factors $f_{\xi_i}(q)$ for each state.
Using these, we can calculate the average form factors and their fluctuations as
\begin{align}
	\erw{f_i(q,t)} &= \sum_{\xi_i} p_{\xi_i}(t) f_{\xi_i}(q) \\
	\erw{|\delta f_i(q,t)|^2} &= \erw{|f_i(q,t)|^2} - |\erw{f_i(q,t)}|^2 = \sum_{\xi_i} p_{\xi_i}(t) |f_{\xi_i}(q)|^2 - |\erw{f_i(q,t)}|^2
	\ .
\end{align}

\end{document}